%% file: paper.tex
\documentclass[11pt]{article}
\usepackage[margin=1.1in]{geometry}
\usepackage{booktabs}
\usepackage{array}
\usepackage{amsmath}
\usepackage{graphicx}
\usepackage[hidelinks]{hyperref}
\usepackage{microtype}
\usepackage{pgfplots}
\pgfplotsset{compat=1.18}
\input{numbers.tex}
\title{More Reader Data or a Better Teacher?\\Updating a Model of Crowd Highlights in Late 2026}
\author{Kazuki Nakayashiki \qquad Keisuke Watanabe\\[2pt]
Glasp Inc.\\
\texttt{\{kazuki,kei\}@glasp.co}}
\date{}

\begin{document}
\maketitle

\begin{abstract}
A model trained on readers' highlights can show a newcomer where earlier readers converged on a page,
even on pages with few readers. A year of stronger language models suggests two upgrades: use them
directly, or use them as teachers. We compare both with a third, more reader data, on one social
highlighting platform, with the comparisons, the selection rules and a sealed set of fresh pages fixed
in advance. On \DevPages{} development pages, seven language models released by October 2026 rank
sentences zero-shot at an average precision of at most \DevAstraMean{} against the sentences half the
readers marked most, and \DevFuseNew{} when fused with reading position, while a 3-billion-parameter
model trained on reader highlights alone reaches \DevHOne{}. Used as the teacher that breaks ties among
readers' counts, the stronger fusion changes a trained model by \DevPilotGainFull{}, whereas going from half to all
of its training pages adds \DevCurveFull{}. A 27-billion-parameter backbone on the reader-only recipe adds
\DevQHvsHOne{} on the development pages; on \TestPages{} fresh pages, read once and too few for a
confirmatory comparison, its difference from the 3-billion model is unresolved, both score above the
fusion, and in an exploratory split their lead over the previous model sits on pages from domains that
their fresh training pages also cover. In this recipe, more reader data improves a trained model where a
stronger tie-breaking teacher yields no material gain, and frontier models used as rankers still trail a
small model trained on readers.
\end{abstract}

\section{Introduction}

Social highlighting platforms let readers see which passages other readers marked. On a page with
few readers there is little to show, and a model of where readers converge (crowd salience) can fill
the gap. Our earlier work found that such a model is learnable from the page alone~\cite{p4}, that
language models agree with one another far more than with readers~\cite{p7}, and that fusing several
frontier models with reading position recovers more of the predictable structure than any single
model~\cite{p8}. The best model we had in August 2026 was a 3-billion-parameter language model
distilled from such a fusion and then trained on readers' highlights: it ranked sentences well above
every model it learned from.

Two months later the frontier had moved. Seven new models were available, and the natural plan is to
ask whether they close the gap zero-shot, and if not, whether they make a better teacher for the
trained model. A third option needs no new model: more reader data. This paper compares the three on the
highlights readers leave in ordinary use, with the comparisons fixed in advance.

\textbf{This paper makes one claim: for predicting where readers converge, reader data still beats
frontier models. As rankers, alone or fused, they trail a small model trained on reader highlights;
and in the recipe we tested, more reader pages improve that model while a stronger frontier teacher,
used to break ties among readers' counts, yields no material gain.} No single new model clears the earlier
reference by the margin we set in advance; the seven fused with reading position do, and still trail
a 3-billion-parameter model trained on reader highlights alone by \DevFuseGapHOne{} average precision.
That stronger fusion, used as the teacher that breaks ties, moves the trained model by \DevPilotGain{},
while going from half to all of the reader pages moves it by \DevCurve{}. A larger backbone scores higher on the pages it
was selected on; on fresh pages its advantage could not be confirmed, and in an exploratory split
the gains of the new models over the previous one sit on familiar domains.

\textbf{Related work.} Distilling a large model's rankings into a small one is established in retrieval,
where a 440-million-parameter student of ChatGPT's rankings outperformed a 3-billion-parameter
supervised ranker~\cite{sun2023}, in the tradition of knowledge distillation~\cite{hinton2015}, and
a language model has outperformed crowd workers on several annotation tasks~\cite{gilardi2023}. Readers'
highlights carry information about their comprehension and interests~\cite{winchell2020}, and in our
earlier work, asking a model which sentences a reader would highlight agreed with readers less than
asking it which sentences to keep~\cite{p6}. What we add is a measured comparison, on naturally
accumulated reader behaviour, of the two levers a platform can pull: a stronger teacher, or more of its
own readers' data.

\section{Setting}

\textbf{Pages, readers and labels.} We use public highlights on one social highlighting platform.
A page's text is split into sentences and each reader's highlights are mapped onto them; a page enters
an evaluation set with at least 6 readers and 30 sentences. Following our earlier protocol~\cite{p8},
the readers of a page are split in half 40 times; on each split the sentences half B marked most
(up to $\mathrm{round}(0.15N)$ of a page's $N$ sentences, only sentences half B marked, ties broken at
random) are the positives, every model's ranking of the page's sentences is scored
by average precision (AP) against them, and half A's counts give a split-half reference (how well half
the readers predict the other half). A page's AP is the mean
over its splits, and a set's AP the mean over its pages. Differences are paired by page with 95\%
intervals from a bootstrap that resamples domains (4,000 draws)~\cite{cameron2008}; they reflect the
sampling of pages, not variation between training runs. Families of comparisons are
Holm-adjusted~\cite{holm1979}.

\textbf{Sets.} The \emph{development} set is our earlier 217-page evaluation set, re-extracted under
the rules below; \DevPages{} pages over \DevDomains{} domains pass the gates. The \emph{training}
pages are the 819 pages of the August model with their readers re-extracted under the same rules
(801 keep at least one reader). \emph{Fresh} pages are pages that never entered any earlier training
or evaluation set, by page identity, URL and text near-duplication. The fresh pages with 3--5 readers
form a second training set (\TrainXPages{} pages); those with at least 6 readers and fetchable public
text form the \emph{test} set (\TestPages{} pages over \TestDomains{} domains). Near-duplicates of
test pages are kept out of the second training set. The test set was sealed when it was written and
read once, after the candidates were frozen. Only highlights public under the platform's own rule, recorded before a
fixed cutoff, were read.

\textbf{What was fixed in advance.} The comparisons, the margins that decide each of them, the rules
that choose the candidates for the test set and the three comparisons on it were registered before
any of them was run. At the test set's reader depth the smallest difference it could detect
(\PowerMdd{}) exceeds half the effect it was meant to confirm (\PowerHalf{}), a condition we fixed in
advance, so every number on the test set is reported as a description, not a confirmatory test.

\section{Frontier models zero-shot}

Seven models released by October 2026 (\texttt{claude-opus-5-5}, \texttt{claude-fable-5-1} and
\texttt{claude-haiku-5-5} at low reasoning effort, \texttt{gpt-6.1-sol} and \texttt{gpt-6-astra} at low
effort, \texttt{gpt-6-luna} without reasoning, \texttt{gemini-3.8-flash} at its default) each ranked
every sentence of a page, given the numbered sentences and asked for the most important first; we
drew two samples per page. A ranking becomes scores by Borda count~\cite{aslam2001}, and a fusion sums
the members' normalised Borda scores with reading position, as in~\cite{p8}. The references are the
cached July 2026 draw of \texttt{gemini-3.1-pro-preview}, the best single model then~\cite{p8}, and
the earlier fusion of five models with position~\cite{p8}.

\begin{table}[t]
\centering\small
\begin{tabular}{lrr}
\toprule
first sample, against \texttt{gemini-3.1-pro-preview} & difference [95\%] & Holm $p$ \\
\midrule
\texttt{gpt-6-astra} (low) & \RtAstraFull & \RtAstraHolm \\
\texttt{gpt-6-astra} (high) & \RtAstraHighFull & \RtAstraHighHolm \\
\texttt{gpt-6.1-sol} & \RtSolFull & \RtSolHolm \\
\texttt{claude-opus-5-5} & \RtOpusFull & \RtOpusHolm \\
\texttt{claude-fable-5-1} & \RtFableFull & \RtFableHolm \\
\texttt{gemini-3.8-flash} & \RtFlashFull & \RtFlashHolm \\
\texttt{claude-haiku-5-5} & \RtHaikuFull & \RtHaikuHolm \\
\texttt{gpt-6-luna} & \RtLunaFull & \RtLunaHolm \\
\midrule
the seven fused with position & \RtFusionFull & \RtFusionHolm \\
\quad against the earlier five-model fusion & \RtFusionOldFull & \\
\bottomrule
\end{tabular}
\caption{Zero-shot ranking on the development set (\RtPages{} pages, one scoring run): each model's
first sample against the cached reference. Holm $p$ is the Holm-adjusted permutation $p$ of the
difference from zero, over the nine rows. The registered rule: a model or the new fusion passes when
its point estimate reaches the margin ($+0.02$ over the reference; for the new fusion also $+0.015$
over the earlier fusion) and its 95\% interval lies above zero, with a Holm $p$ below 0.05 for the comparison with the
reference. The margin bounds the point estimate, not the interval.}
\label{tab:zeroshot}
\end{table}

No single model clears the margin of $+0.02$ we set in advance (Table~\ref{tab:zeroshot}); the nearest,
\texttt{gpt-6-astra}, gains \RtAstra{}, and high reasoning effort scores no higher. Two samples of the same
model agree at a median per-page Spearman correlation between \AgreeMin{} and \AgreeMax{}, and the
mean of two samples adds at most \DevMeanGainMax{} AP. Fusion does what it did before~\cite{p8}: the
seven with position beat the reference by \RtFusion{} and the earlier fusion by \RtFusionOld{}, both
resolved. The fusion reaches \RtFuseAp{} AP in this run.

In an exploratory check added after the test set was read, on the development set, a prompt aimed at
readers does not help. Asked instead for the sentences a typical reader would highlight (the
instruction of~\cite{p6}, adapted to rank every sentence; Appendix~\ref{app:repro}),
\texttt{gpt-6-astra} changes by \PromptAstraFull{} and \texttt{claude-opus-5-5} by \PromptOpusFull{}
(first samples, in a scoring run of their own), as \texttt{gpt-5.4} did in~\cite{p6}.

\section{Training on readers}

\textbf{Model.} A pretrained language model reads the numbered sentences of a page (up to 8,192
tokens) and a two-layer head scores each sentence from the hidden state at its last token; the base is
quantised to 4 bits and adapted with LoRA~\cite{hu2022,dettmers2023}. The target ranks sentences by how
many readers marked them, and ties, which dominate on pages with few readers, are broken either by a
teacher's ranking or by position (earlier first). The loss is an equal mix of squared error on the
standardised target and a listwise cross-entropy at temperature 0.1. The \emph{August model} is
SmolLM3-3B first distilled from the earlier five-model fusion and then trained on the 819 training
pages with that fusion breaking ties; it comes from an unpublished comparison of backbones and is
reported here for the first time. Every model trained for this paper is trained twice with different seeds
and reported as the mean of the two; the August model and the other backbones of its comparison are
single earlier runs.

\begin{table}[t]
\centering\small
\begin{tabular}{>{\raggedright\arraybackslash}p{4.4cm}>{\raggedright\arraybackslash}p{7.0cm}r}
\toprule
model & training & AP \\
\midrule
27B, readers only & Qwen3.8-27B from base; fresh pages, then the 819; position breaks ties & \DevQH \\
27B, teacher breaks ties & Qwen3.8-27B from base; the 819; the five-model fusion breaks ties & \DevQT \\
August model & SmolLM3-3B, distilled start; the 819 (August readers) & \DevAug \\
August recipe, re-run & the same recipe trained again here, on the same GPU as every arm & \DevCtl \\
new teacher & distilled start, 100\%, the seven-model fusion breaking ties & \DevPilot \\
3B, readers only & SmolLM3-3B from base; fresh pages, then the 819 & \DevHOne \\
distilled start, 100\% & SmolLM3-3B, distilled start; the 819, refreshed readers & \DevPHundred \\
3B, teacher breaks ties & SmolLM3-3B from base; the 819; the five-model fusion breaks ties & \DevSmolT \\
3B, readers only, 819 & SmolLM3-3B from base; the 819; position breaks ties & \DevHZero \\
distilled start, 50\% / 25\% & as 100\% on half / a quarter of the pages & \DevPFifty{} / \DevPQuarter \\
\midrule
the seven fused with position & zero-shot & \DevFuseNew \\
the earlier five-model fusion & zero-shot & \DevFuseOld \\
best single model & \texttt{gpt-6-astra}, mean of two samples & \DevAstraMean \\
S3 & Qwen3-8B distilled from the five-model fusion~\cite{p8} & \DevSThree \\
reading position & first sentences first & \DevLead \\
split-half reference & half the readers predicting the other half & \DevCeiling \\
\bottomrule
\end{tabular}
\caption{Development set, \DevPages{} pages, all models in one scoring run; the differences in the
text are computed in this run from unrounded per-page values. ``Fresh pages'' are the
\TrainXPages{} fresh pages with 3--5 readers.}
\label{tab:dev}
\end{table}

\begin{figure}[t]
\centering
\begin{tikzpicture}
\begin{axis}[width=0.7\linewidth, height=4.4cm, xmode=log, log basis x=2, xmin=0.21, xmax=1.3,
  xtick={0.25,0.5,1}, xticklabels={25\%,50\%,100\%}, xlabel={share of the 819 training pages},
  ylabel={AP, development set}, ymin=0.37, ymax=0.445, ytick={0.38,0.40,0.42,0.44},
  yticklabel style={/pgf/number format/fixed, /pgf/number format/fixed zerofill, /pgf/number format/precision=2},
  tick label style={font=\scriptsize}, label style={font=\small},
  legend style={font=\scriptsize, at={(0.5,-0.36)}, anchor=north, draw=none, legend columns=2,
    /tikz/every even column/.append style={column sep=8pt}}, legend cell align=left]
\addplot[mark=*, thick] coordinates {(0.25,\DevPQuarter) (0.5,\DevPFifty) (1,\DevPHundred)};
\addlegendentry{distilled start, earlier teacher}
\addplot[only marks, mark=triangle*, mark size=3pt] coordinates {(1.12,\DevPilot)};
\addlegendentry{all pages, stronger teacher}
\addplot[dashed, domain=0.21:1.3, samples=2] {\DevAug};
\addlegendentry{August model}
\addplot[dotted, thick, domain=0.21:1.3, samples=2] {\DevFuseNew};
\addlegendentry{seven frontier models fused, zero-shot}
\end{axis}
\end{tikzpicture}
\caption{More pages against a stronger tie-breaking teacher on the development set (the run of Table~\ref{tab:dev}): the
distilled start trained on a quarter, half and all of the 819 training pages with the earlier teacher
breaking ties, and on all of them with the stronger teacher (the triangle, drawn right of 100\% for
legibility).}
\label{fig:curve}
\end{figure}
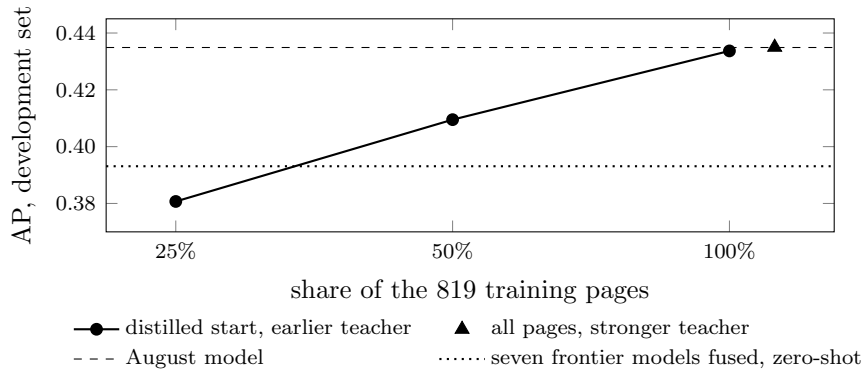

\textbf{More pages help; refreshed labels and a stronger tie-breaking teacher add little.} Table~\ref{tab:dev} gives
every arm, and every comparison on the development set below is computed in that one scoring run (the
registered decisions and the freeze of Section~\ref{sec:fresh} were taken in earlier runs and hold in this one). On nested quarters, halves and all of the 819 pages, the distilled start gains \DevCurveFull{}
from the half to the whole (Figure~\ref{fig:curve}). Re-extracting the readers of the same pages changes the trained model by
\DevRefreshFull{} against the August recipe re-run, a change that includes losing the 18 pages left
without a reader. The seven-model fusion, which ranks pages \DevFusionGap{} better than the earlier
fusion in this run, was collected for all 819 training pages (each model's first sample, fused with position) and
used to break ties in an otherwise identical run: the model moves by \DevPilotGainFull{}, below the
$+0.005$ we required, and by \PilotSeedZeroFull{} and \PilotSeedOneFull{} with either seed alone. The swap
changes the target's order for \PairsDiffer{} of the sentence pairs on these pages: the target orders
sentences by readers' counts, so a teacher orders only pairs with equal counts (\TiedShare{} of them), and
the two teachers order \TiedAlike{} of those alike; \ChangedUnmarked{} of the pairs that change are of
two sentences no reader marked. A teacher is not useless: from the base model on the 819 pages, breaking ties
with the five-model fusion instead of position adds \DevTeacherFromBaseFull{}. But once a teacher
breaks the ties, a stronger one adds little next to more reader pages.

\textbf{A model trained on readers alone comes close to the distilled one.} Trained from the base model on
reader highlights with position breaking ties, SmolLM3-3B reaches \DevHZero{} on the 819 pages and
\DevHOne{} when the \TrainXPages{} fresh pages come first (\DevHOnevsHZeroFull{}), a point estimate
close to the August model's (\DevHOnevsAugFull{}; the difference is unresolved). No output of a frontier model enters it.
An exploratory control on the development set, added after the test set was read, shows that the gain
comes from the pages,
not from the 6,320 steps that come with them: trained as long on the 819 pages alone, the model changes
by \EqStepsFull{} against the shorter run, and the fresh pages lead that control by \EqFreshFull{} (a
scoring run of their own).

\textbf{A larger backbone scores higher here.} The same recipes on Qwen3.8-27B, of another family and a
newer generation than SmolLM3-3B, score \DevQTvsSmolTFull{}
above SmolLM3-3B with the five-model fusion breaking ties, and \DevQHvsHOneFull{} above it with readers only;
the 27B reader-only model reaches \DevQH{}, against the August model's \DevAug{} (\DevQHvsAugFull{}). These are the
pages on which the candidates were selected.

\begin{table}[t]
\centering\small
\begin{tabular}{lrr}
\toprule
backbone & August recipe & readers only \\
\midrule
ModernBERT-large (0.4B, encoder) & \LadMbert & --- \\
Qwen3-0.6B & \LadSix & \SmallSix \\
Qwen3-1.7B & \LadOneSeven & \SmallOneSeven \\
SmolLM3-3B & \DevAug & \DevHOne \\
Qwen3-4B & \LadFour & --- \\
Qwen3-8B & \LadEight & --- \\
Qwen3.8-27B & --- & \DevQH \\
\midrule
the seven fused with position (zero-shot) & \multicolumn{2}{c}{\DevFuseNew} \\
\bottomrule
\end{tabular}
\caption{Backbone size on the development set (exploratory; AP in the run of Table~\ref{tab:dev}).
\emph{August recipe}: distilled from the five-model fusion, then trained on the 819 pages with that
fusion breaking ties; one run each, from the comparison the August model (the SmolLM3-3B row) came from.
\emph{Readers only}: the recipe of the 3B reader-only model, mean of two seeds; the 0.6B and 1.7B models
were trained after the test set was read and were not scored on it.}
\label{tab:size}
\end{table}

\textbf{Smaller backbones (exploratory).} After the test set was read, we trained the reader-only recipe
on Qwen3-0.6B and Qwen3-1.7B and scored them in the same run, next to the other backbones of the
comparison the August model came from (Table~\ref{tab:size}). At both sizes the reader-only model's
point estimate is close to that of the distilled model of the same size (\SizeSixGainFull{} at 0.6B,
\SizeOneSevenGainFull{} at 1.7B; unresolved), as at 3B. The score rises with size: the 0.6B reader-only
model trails the 3B one (\SizeSixVsHOneFull{}), yet its point estimate still exceeds the seven-model
fusion's (\SizeSixVsFusionFull{}), as does that of the 0.4B encoder on the August recipe
(\SizeMbertVsFusionFull{}). Neither new model was a candidate, and neither was scored on the test set.

\section{Fresh pages}
\label{sec:fresh}

We froze two candidates by the registered rules (the best 3B reader-only model and the best
model overall, the 27B reader-only model, which is also the larger size) and the best zero-shot
configuration (the seven-model fusion, collected on the test pages with one sample per model), and
scored them once on the sealed test set with the August model and S3. The checks of the scoring
held: a random ranking sits at its expectation (\TestAssayRandomFull{}), a permuted ranking at random
(\TestAssayPermFull{}), and S3 above position (\TestAssayLeadFull{}).

\begin{table}[t]
\centering\small
\begin{tabular}{lr@{\hspace{2.5em}}lr}
\toprule
model & AP & comparison & difference [98.3\%] \\
\midrule
3B, readers only & \TestHOne & C1: August model $-$ S3 & \TestCOneFull \\
27B, readers only & \TestQH & C2: 27B $-$ August model & \TestCTwoFull \\
August model & \TestAug & C3: 27B $-$ the fusion & \TestCThreeFull \\
the seven fused with position & \TestFuse & & \\
S3 & \TestSThree & & \\
reading position & \TestLead & & \\
random (expected) & \TestRandom & & \\
\bottomrule
\end{tabular}
\caption{The sealed test set: \TestPages{} fresh pages over \TestDomains{} domains, one scoring run,
intervals Bonferroni-adjusted over the three registered comparisons. Descriptive: the set is smaller
than a confirmatory comparison needed.}
\label{tab:test}
\end{table}

The August model keeps its lead over S3 on fresh pages (C1 in Table~\ref{tab:test}), on pages with
6--9 readers (\TestCOneShallowFull{}, \TestShallowPages{} pages) and with 10--19
(\TestCOneMidFull{}, \TestMidPages{} pages). The 27B model leads the seven-model fusion (C3), and so does
the 3B (\TestHOnevsFuseFull{}, an unregistered comparison). The 27B's
lead over the August model (C2) does not exclude zero, and its difference from the
3B is unresolved (\TestQHvsHOneFull{}, an unregistered comparison). The split-half reference (\TestCeiling{}) sits below
the two reader-only candidates because most test pages have 6--7 readers, so three readers predict three
or four.

\textbf{Exploratory, after the look.} \FamTestPages{} of the \TestPages{} test pages come from domains
that also hold a fresh training page. On those, the 3B and 27B reader-only models lead the August model
by \FamTestHOneFull{} and \FamTestQHFull{}; on the other \UnfamTestPages{} pages, from
\UnfamTestDomains{} domains (\UnfamTestNeither{} of them absent from the 819 training pages too), they
trail it by \UnfamTestHOneFull{} and \UnfamTestQHFull{} (95\% intervals). On the development set the
3B's and the 27B's differences from the August model are \FamDevHOneFull{} and \FamDevQHFull{} on the
\FamDevPages{} pages of such domains, and \UnfamDevHOneFull{} and \UnfamDevQHFull{} on the other
\UnfamDevPages{}. Eleven pages
cannot settle it, but the gain of the new models may be coverage of the domains a platform's readers
keep reading rather than a better general ranker.

\section{Discussion}

Frontier models rank sentences by what a model finds important, and readers converge on something
else~\cite{p7}. A newer generation narrows that gap only a little: the best single model of October
2026 stays \DevBestSingleGap{} AP below a 3B model trained on readers, and fusing seven of them still
leaves \DevFuseGapHOne{}. The trained model's advantage comes from the readers, and the parts of its
recipe that rely on a frontier model matter less than they seemed. A teacher that breaks ties helps a
model trained on few reader pages, but a stronger one added little once a teacher was there, and with \TrainXPages{}
more reader pages a model trained from its base model on readers alone reaches the distilled model's
point estimate. For
a platform this is the useful direction: the improvement comes from data it already collects, needs
no model provider's outputs in training, and stays with a small model (Table~\ref{tab:size}).

The larger backbone is the open question. On the development set it adds as much as doubling the
pages did; on fresh pages its advantage could not be confirmed, and the exploratory split points at domain coverage as a
common source of the new models' gains there. A test on pages from domains that no training page
covers would separate a better ranker from a familiar one.

\textbf{Limits.} One platform, public pages whose text could be fetched, and Latin-script pages only
in the fresh training and test sets. The test set is small, was read as a description, and tests the
frozen candidates, not the development-set comparisons behind them. The stronger teacher replaced only
the tie-breaker: the distilled start kept the earlier teacher, so distilling anew from the stronger one,
or giving the fresh pages to the distilled model, was not tested. The zero-shot models are specific
versions at specific settings, which date quickly; a reader-prediction instruction lowered the two we
tried, but other prompts, examples or fine-tuning might narrow the gap. The trained models were
selected on the development set, so their numbers there favour them.

\section{Conclusion}

Seven language models released by October 2026 do not close the gap to a small model trained on
reader highlights, alone or fused, and the stronger fusion, as the teacher that breaks ties, did not
improve it by the margin we required. More reader pages did, and a model trained on reader highlights
alone now reaches a point estimate close to one distilled from frontier models. On this platform and in
the recipe we tested, the data that improves a model of where readers converge is the readers'.

\section*{Ethics and data}
This is a retrospective, non-interventional secondary analysis of service records collected under the
platform's Terms of Service and Privacy Policy, which name research among the purposes of processing.
The study read production records read-only, kept only highlight documents public under the
platform's own rule (public is the default), and wrote records without account identifiers or reader
keys. Page text came from the platform's stored copies or, for fresh pages, from the public web without
credentials; model providers received public page text only, and the training service received page
text and highlight positions keyed by ordinals. Record-level files are not released, and every reported
result is an aggregate over pages with at least six readers. The operating company is not a research
institution and has no institutional review board; no institutional review board reviewed this study,
and no formal ethics approval was sought. Both authors are affiliated with the platform's operator.

\section*{AI assistance}
The study design, the extraction, training and analysis code, the statistics, the tables and a first
draft of this paper were produced with an AI coding agent (Anthropic's Claude) under the authors'
direction. OpenAI's Codex and a separate Claude instance audited the design, the code and the
write-ups. The authors take full responsibility for the content.

\section*{Data and code availability}
Record-level data cannot be released. The ancillary files contain the pre-registration with its
amendments, the results write-up, every aggregate artifact, and the code that builds the prompts,
scores the rankings and builds the training targets. Every number in this paper regenerates from those
artifacts (\texttt{make-numbers.py}), and \texttt{check-claims.py} checks the paper's listed qualitative
comparisons against them; retraining a model needs the record-level data, so the training and scoring
code is there as specification.

\appendix
\section{Reproduction details}
\label{app:repro}

\textbf{Zero-shot prompt.} Every model received the system message ``You compress documents by
selecting the most important sentences. Output only what is asked for.'' and a user message that
lists the page's sentences, one per line as ``[$i$] sentence'' after ``Sentences from one article, in
order:'', and ends ``You must compress this document by keeping only the most important sentences.
Rank ALL $N$ sentence numbers from most important to least important. Return ONLY a JSON array of
every sentence number, most important first.'' It is the prompt of~\cite{p8}, kept byte-identical so
that the new models compare with its cached draws. The reader-prediction instruction (Section 3) takes
the system message of~\cite{p6}, ``You select the sentences a reader would most likely highlight or
underline while reading. Readers highlight memorable, surprising, quotable, or personally striking
passages --- not merely topic sentences or definitions. Output only JSON.'', and after the same list of
sentences asks ``Rank ALL $N$ sentence numbers from the one a typical reader would most likely highlight
to the one least likely. Return ONLY a JSON array of every sentence number, most likely first.''

\textbf{Models and settings} (collected on 2026-10-08 and 2026-10-09). \texttt{gpt-6.1-sol} and
\texttt{gpt-6-astra} through OpenAI's API at reasoning effort low (\texttt{gpt-6-astra} also at high,
one sample); \texttt{gpt-6-luna} at effort none and temperature 0; \texttt{anthropic/claude-opus-5.5},
\texttt{anthropic/claude-fable-5.1} and \texttt{anthropic/claude-haiku-5.5} at effort low, pinned to
Anthropic, and \texttt{google/gemini-3.8-flash} at its default, through OpenRouter. No other
temperature was sent; the output cap was 32,000 tokens (64,000 at high effort). Batch and
synchronous calls carried identical requests. A ranking that leaves sentences out keeps the ranked
ones above the rest; a refusal, or a filtered, truncated or unparseable answer, is scored as a tie of
all sentences (\ZsFailures{} of the \ZsCells{} first samples on the development set).

\textbf{Training.} The base model is quantised to 4-bit NF4 and adapted with LoRA of rank 16
($\alpha = 32$, dropout 0.05) on the attention and MLP projections~\cite{hu2022,dettmers2023}; the head
is two linear layers with 256 hidden units and GELU. Each sentence is written as ``$i$: sentence'' and a
page is cut at 8,192 tokens; sentences past the cut score below every scored one, in document order.
The target of sentence $i$ is the rank fraction, within the page, of $c_i + t_i$, where $c_i$ is
its reader count and $t_i \in [0, 1)$ the rank fraction of the teacher's score or of position, so the
tie-breaker orders sentences only within the same count. The loss is $0.5$ times the squared error on the
standardised target plus $0.5$ times the Kullback--Leibler divergence between the softmax of the target
and that of the prediction at temperature 0.1. AdamW at $2\times10^{-5}$ when continuing the distilled
start and $10^{-4}$ for a fresh adapter; gradients are accumulated over 8 pages. A run takes a fixed
budget of 2,457 page steps on the training pages (306 updates, 304 on the quarter of the pages;
August's three passes over its 819 pages), after 6,320 steps on the fresh pages for the reader-only recipe (two passes, 790 updates). The page order is shuffled once per seed, the model
is the last step (validation is recorded, not used to choose), and every arm trained here has two seeds,
on one NVIDIA H100. Qwen3-0.6B and Qwen3-1.7B (Table~\ref{tab:size}) use the settings of the 3B reader-only
model. The equal-steps control (Section 4) is the 3B reader-only model's command with the first stage
reading the 819 training pages instead of the fresh ones. The August-recipe rows of Table~\ref{tab:size} are the stored sentence scores of the earlier
comparison's runs, one each, made with the August model's two stages; the 8B's distilled start is S3,
and ModernBERT-large, an encoder, was fully fine-tuned, from a distilled start trained for eight passes
at $10^{-4}$.

\textbf{Pages and readers.} A reader's highlights are mapped onto the sentences they cover, and a
sentence's count is the number of readers who marked it. Of the 819 training pages, 18 kept no reader
whose highlights still anchor to the stored text after re-extraction, so the runs on the re-extracted
readers use the other 801 (a subset, the part of them it holds), while the August recipe re-run uses
the August readers of all 819; the teachers were collected for all 819.

\textbf{Registration.} The pre-registration was fixed on 2026-10-08, before any record was read; its
dated amendments, the candidates frozen on 2026-10-09 before the test set was opened, and the results
write-up are in the ancillary files, with the code that builds the prompts, scores the rankings and
regenerates every number.

\end{document}

%% file: numbers.tex
\newcommand{\AgreeMax}{0.97}
\newcommand{\AgreeMin}{0.70}

\newcommand{\ChangedUnmarked}{86\%}

\newcommand{\DevAstraMean}{0.3744}
\newcommand{\DevAug}{0.4349}
\newcommand{\DevBestSingleGap}{0.060}
\newcommand{\DevCeiling}{0.5136}
\newcommand{\DevCtl}{0.4345}
\newcommand{\DevCurve}{+0.0242}

\newcommand{\DevCurveFull}{+0.0242 [+0.0179, +0.0301]}

\newcommand{\DevDomains}{84}

\newcommand{\DevFuseGapHOne}{0.042}

\newcommand{\DevFuseNew}{0.3931}
\newcommand{\DevFuseOld}{0.3765}
\newcommand{\DevFusionGap}{+0.0166}

\newcommand{\DevHOne}{0.4346}

\newcommand{\DevHOnevsAugFull}{\ensuremath{-}0.0003 [\ensuremath{-}0.0219, +0.0287]}

\newcommand{\DevHOnevsHZeroFull}{+0.0245 [+0.0100, +0.0472]}

\newcommand{\DevHZero}{0.4101}

\newcommand{\DevLead}{0.2509}

\newcommand{\DevMeanGainMax}{0.006}

\newcommand{\DevPFifty}{0.4095}
\newcommand{\DevPHundred}{0.4337}
\newcommand{\DevPQuarter}{0.3807}
\newcommand{\DevPages}{215}
\newcommand{\DevPilot}{0.4350}
\newcommand{\DevPilotGain}{+0.0013}

\newcommand{\DevPilotGainFull}{+0.0013 [\ensuremath{-}0.0012, +0.0040]}

\newcommand{\DevQH}{0.4572}

\newcommand{\DevQHvsAugFull}{+0.0222 [\ensuremath{-}0.0003, +0.0493]}

\newcommand{\DevQHvsHOne}{+0.0225}

\newcommand{\DevQHvsHOneFull}{+0.0225 [+0.0155, +0.0295]}

\newcommand{\DevQT}{0.4465}

\newcommand{\DevQTvsSmolTFull}{+0.0225 [+0.0153, +0.0308]}

\newcommand{\DevRefreshFull}{\ensuremath{-}0.0008 [\ensuremath{-}0.0030, +0.0015]}

\newcommand{\DevSThree}{0.3641}
\newcommand{\DevSmolT}{0.4240}

\newcommand{\DevTeacherFromBaseFull}{+0.0139 [+0.0037, +0.0235]}

\newcommand{\EqFreshFull}{+0.0446 [+0.0299, +0.0699]}

\newcommand{\EqStepsFull}{\ensuremath{-}0.0201 [\ensuremath{-}0.0273, \ensuremath{-}0.0134]}

\newcommand{\FamDevHOneFull}{+0.0103 [\ensuremath{-}0.0241, +0.0653]}

\newcommand{\FamDevPages}{125}

\newcommand{\FamDevQHFull}{+0.0337 [\ensuremath{-}0.0010, +0.0823]}

\newcommand{\FamTestHOneFull}{+0.0496 [+0.0020, +0.1377]}

\newcommand{\FamTestPages}{120}

\newcommand{\FamTestQHFull}{+0.0445 [+0.0066, +0.1058]}

\newcommand{\LadEight}{0.4371}
\newcommand{\LadFour}{0.4318}
\newcommand{\LadMbert}{0.4043}
\newcommand{\LadOneSeven}{0.4234}
\newcommand{\LadSix}{0.4138}
\newcommand{\PairsDiffer}{8\%}

\newcommand{\PilotSeedOneFull}{+0.0013 [\ensuremath{-}0.0012, +0.0039]}

\newcommand{\PilotSeedZeroFull}{+0.0013 [\ensuremath{-}0.0017, +0.0044]}

\newcommand{\PowerHalf}{0.0179}
\newcommand{\PowerMdd}{0.0247}

\newcommand{\PromptAstraFull}{\ensuremath{-}0.0412 [\ensuremath{-}0.0585, \ensuremath{-}0.0282]}

\newcommand{\PromptOpusFull}{\ensuremath{-}0.0363 [\ensuremath{-}0.0582, \ensuremath{-}0.0187]}

\newcommand{\RtAstra}{+0.0159}

\newcommand{\RtAstraFull}{+0.0159 [+0.0050, +0.0261]}

\newcommand{\RtAstraHighFull}{+0.0138 [+0.0045, +0.0229]}

\newcommand{\RtAstraHighHolm}{0.050}

\newcommand{\RtAstraHolm}{0.065}

\newcommand{\RtFableFull}{+0.0073 [\ensuremath{-}0.0060, +0.0193]}

\newcommand{\RtFableHolm}{0.582}

\newcommand{\RtFlashFull}{+0.0015 [\ensuremath{-}0.0082, +0.0097]}

\newcommand{\RtFlashHolm}{0.741}

\newcommand{\RtFuseAp}{0.3909}
\newcommand{\RtFusion}{+0.0345}

\newcommand{\RtFusionFull}{+0.0345 [+0.0228, +0.0450]}

\newcommand{\RtFusionHolm}{0.001}

\newcommand{\RtFusionOld}{+0.0154}

\newcommand{\RtFusionOldFull}{+0.0154 [+0.0072, +0.0238]}

\newcommand{\RtHaikuFull}{\ensuremath{-}0.0180 [\ensuremath{-}0.0320, \ensuremath{-}0.0056]}

\newcommand{\RtHaikuHolm}{0.050}

\newcommand{\RtLunaFull}{\ensuremath{-}0.0248 [\ensuremath{-}0.0370, \ensuremath{-}0.0109]}

\newcommand{\RtLunaHolm}{0.015}

\newcommand{\RtOpusFull}{+0.0120 [\ensuremath{-}0.0019, +0.0240]}

\newcommand{\RtOpusHolm}{0.336}

\newcommand{\RtPages}{215}

\newcommand{\RtSolFull}{+0.0131 [+0.0029, +0.0227]}

\newcommand{\RtSolHolm}{0.094}

\newcommand{\SizeMbertVsFusionFull}{+0.0112 [\ensuremath{-}0.0080, +0.0313]}

\newcommand{\SizeOneSevenGainFull}{+0.0034 [\ensuremath{-}0.0165, +0.0323]}

\newcommand{\SizeSixGainFull}{+0.0020 [\ensuremath{-}0.0180, +0.0307]}

\newcommand{\SizeSixVsFusionFull}{+0.0227 [\ensuremath{-}0.0010, +0.0512]}

\newcommand{\SizeSixVsHOneFull}{\ensuremath{-}0.0189 [\ensuremath{-}0.0252, \ensuremath{-}0.0107]}

\newcommand{\SmallOneSeven}{0.4268}
\newcommand{\SmallSix}{0.4158}

\newcommand{\TestAssayLeadFull}{+0.0748 [+0.0550, +0.0927]}

\newcommand{\TestAssayPermFull}{\ensuremath{-}0.0006 [\ensuremath{-}0.0137, +0.0078]}

\newcommand{\TestAssayRandomFull}{+0.0042 [\ensuremath{-}0.00003, +0.0115]}

\newcommand{\TestAug}{0.3255}

\newcommand{\TestCOneFull}{+0.0420 [+0.0190, +0.0671]}

\newcommand{\TestCOneMidFull}{+0.0429 [+0.0021, +0.0862]}

\newcommand{\TestCOneShallowFull}{+0.0422 [+0.0194, +0.0707]}

\newcommand{\TestCThreeFull}{+0.0603 [+0.0164, +0.1276]}

\newcommand{\TestCTwoFull}{+0.0377 [\ensuremath{-}0.0036, +0.1120]}

\newcommand{\TestCeiling}{0.3275}

\newcommand{\TestDomains}{42}
\newcommand{\TestFuse}{0.3029}

\newcommand{\TestHOne}{0.3663}

\newcommand{\TestHOnevsFuseFull}{+0.0634 [+0.0095, +0.1632]}

\newcommand{\TestLead}{0.2086}
\newcommand{\TestMidPages}{22}
\newcommand{\TestPages}{131}
\newcommand{\TestQH}{0.3632}

\newcommand{\TestQHvsHOneFull}{\ensuremath{-}0.0031 [\ensuremath{-}0.0372, +0.0160]}

\newcommand{\TestRandom}{0.1702}
\newcommand{\TestSThree}{0.2834}
\newcommand{\TestShallowPages}{107}
\newcommand{\TiedAlike}{85\%}
\newcommand{\TiedShare}{54\%}
\newcommand{\TrainXPages}{3{,}160}

\newcommand{\UnfamDevHOneFull}{\ensuremath{-}0.0150 [\ensuremath{-}0.0342, \ensuremath{-}0.0002]}

\newcommand{\UnfamDevPages}{90}

\newcommand{\UnfamDevQHFull}{+0.0062 [\ensuremath{-}0.0162, +0.0236]}

\newcommand{\UnfamTestDomains}{11}

\newcommand{\UnfamTestHOneFull}{\ensuremath{-}0.0545 [\ensuremath{-}0.1095, \ensuremath{-}0.0148]}

\newcommand{\UnfamTestNeither}{10}
\newcommand{\UnfamTestPages}{11}

\newcommand{\UnfamTestQHFull}{\ensuremath{-}0.0362 [\ensuremath{-}0.0804, \ensuremath{-}0.0028]}

\newcommand{\ZsCells}{1{,}505}
\newcommand{\ZsFailures}{7}